\documentclass[
    ,final            
  ]
  {aipproc}

\layoutstyle{8x11double}

\newcommand{\bn}{\bar n}

\newcommand{\nn}{\nonumber} 
\newcommand{\mcdot}{\!\cdot\!}
\newcommand{\nslash}{n\!\!\!\!\slash}

\newcommand{\lesssim}{\raisebox{-.25em}{$\stackrel{\mbox{\small $<$}}{\sim}
$}}

\begin{document}

\title{Factorization and the Soft-Collinear Effective Theory: \\[2pt]
 Color-Suppressed Decays}

\author{Sonny Mantry}{
address={Center for Theoretical Physics, Laboratory for Nuclear Science and
Department of Physics,\\[3pt]
Massachusetts Institute for Technology, Cambridge, MA 02139}
}

\author{Dan Pirjol}{
address={Center for Theoretical Physics, Laboratory for Nuclear Science and
Department of Physics,\\[3pt]
Massachusetts Institute for Technology, Cambridge, MA 02139}
}

\author{Iain W. Stewart\footnote{Plenary talk given by I.S. at the 9'th
  International Conference on B Physics at Hadron Machines (Beauty,
  2003). MIT-CTP 3462}
 $\:$ }{ 
address={Center for Theoretical Physics, Laboratory for Nuclear Science and
    Department of Physics,\\[3pt]
    Massachusetts Institute for Technology, Cambridge, MA 02139} 
}


%

\begin{abstract}
  We discuss the soft-collinear effective theory (SCET) and kinematic expansions
  in $B$-decays, focusing on recent results for color suppressed $B\to D^{(*)}
  X$ decays. In particular we discuss model independent predictions for $\bar
  B^0\to D^{0}\pi^0$ and $\bar B^0\to D^{*0}\pi^0$, and update the comparison
  using new experimental data.  We show why HQET alone is insufficient to give
  these results. SCET predictions are also reviewed for other $B$ and
  $\Lambda_b$ decay channels that are not yet tested by data.

\end{abstract}

\maketitle


\section{Introduction}
  
The soft-collinear effective
theory~\cite{Bauer:2000ew,Bauer:2000yr,Bauer:2001ct,Bauer:2001yt} (SCET)
provides a formalism for systematically investigating processes with both
energetic and soft hadrons based solely on the underlying structure of QCD.
Essentially all known methods for simplifying QCD predictions, without
introducing model dependent assumptions, depend on exploiting hierarchies of
mass scales. For predictions based on SU(3) symmetry we exploit the fact that
$m_{u,d,s}/\Lambda\ll 1$, and expect corrections at the $\sim 30\%$ level.  In
lattice QCD simulations we choose our lattice spacing $a \ll 1/\Lambda$ and
volume $V\gg 1/\Lambda^3$ so that we can focus on non-perturbative effects at
scales $\sim \Lambda$. In SCET we expand in $\Lambda/Q\ll 1$, with the large
momentum of an energetic hadron or jet being $\sim Q$. For $B$ decays
corrections will be at the $\sim 20$--$30\%$ level depending on the energy
scale $Q$.

Most effective theories that we are familiar with are designed to separate the
physics for hard $p_h^2 \simeq Q^2$ and soft $p_s^2 \ll Q^2$ momenta.  Examples
include the electroweak Hamiltonian, chiral perturbation theory, heavy
quark effective theory, and non-relativistic QCD. In SCET we incorporate an
additional possibility, namely energetic hadrons where the constituents have
momenta $p_c^\mu$ nearly collinear to a light-like direction $n^\mu$.  Both the
energetic hadron and its collinear constituents have $\bn\cdot p_c \sim Q$,
where we have made use of light-cone coordinates $(p_c^+,p_c^-,p_c^\perp) =
(n\mcdot p_c,\bn\mcdot p_c,p_c^\perp)$.  The collinear constituents still have
small offshellness $p_c^2\sim p_s^2$.  The process of disentangling the
interactions of hard-collinear-soft particles is known as factorization, and is
simplified by the SCET framework.

Much like any effective theory the basic ingredients of SCET are its field
content, power counting, and symmetries. The Lagrangian and operators, are
organized in a series where only ${\cal L}^{(0)}$ and ${\cal O}^{(0)}$ are
relevant at LO, an additional ${\cal L}^{(1)}$ or ${\cal O}^{(1)}$ is needed at
NLO, etc. The expansion parameter will be $\lambda=\sqrt{\Lambda_{\rm QCD}/Q}$
or $\eta=\Lambda_{\rm QCD}/Q$ depending on whether the collinear fields describe
an energetic jet of hadrons or an individual energetic hadron.  The effective
theory with an expansion in $\lambda$ is called ${\rm SCET}_{\rm I}$, while the
one with an expansion in $\eta$ is called ${\rm SCET}_{\rm II}$. In processes
such as color-suppressed decays the separation of scales is $Q^2 \gg Q\Lambda
\gg \Lambda^2$ and the chain QCD--${\rm SCET}_{\rm I}$--${\rm SCET}_{\rm II}$
proves to be useful. The intermediate theory ${\rm SCET}_{\rm I}$ provides the
dynamics to rearrange soft and collinear quark lines so that they can end up in
soft and energetic hadrons. The final theory ${\rm SCET}_{\rm II}$ describes the
universal low energy hadronic matrix elements. In the case of color-suppressed
decays $B\to D^{(*)}M$ these are light-cone distribution functions $\phi_M(x)$
where $M=\pi,\rho,K,$ or $K^*$ and two generalized parton distribution functions
$S^{(0,8)}(k_1^+,k_2^+)$ for the $B\to D^{(*)}$ transition.

\section{Color-Suppressed Decays and SCET}

\begin{figure}
  \includegraphics[height=.1\textheight]{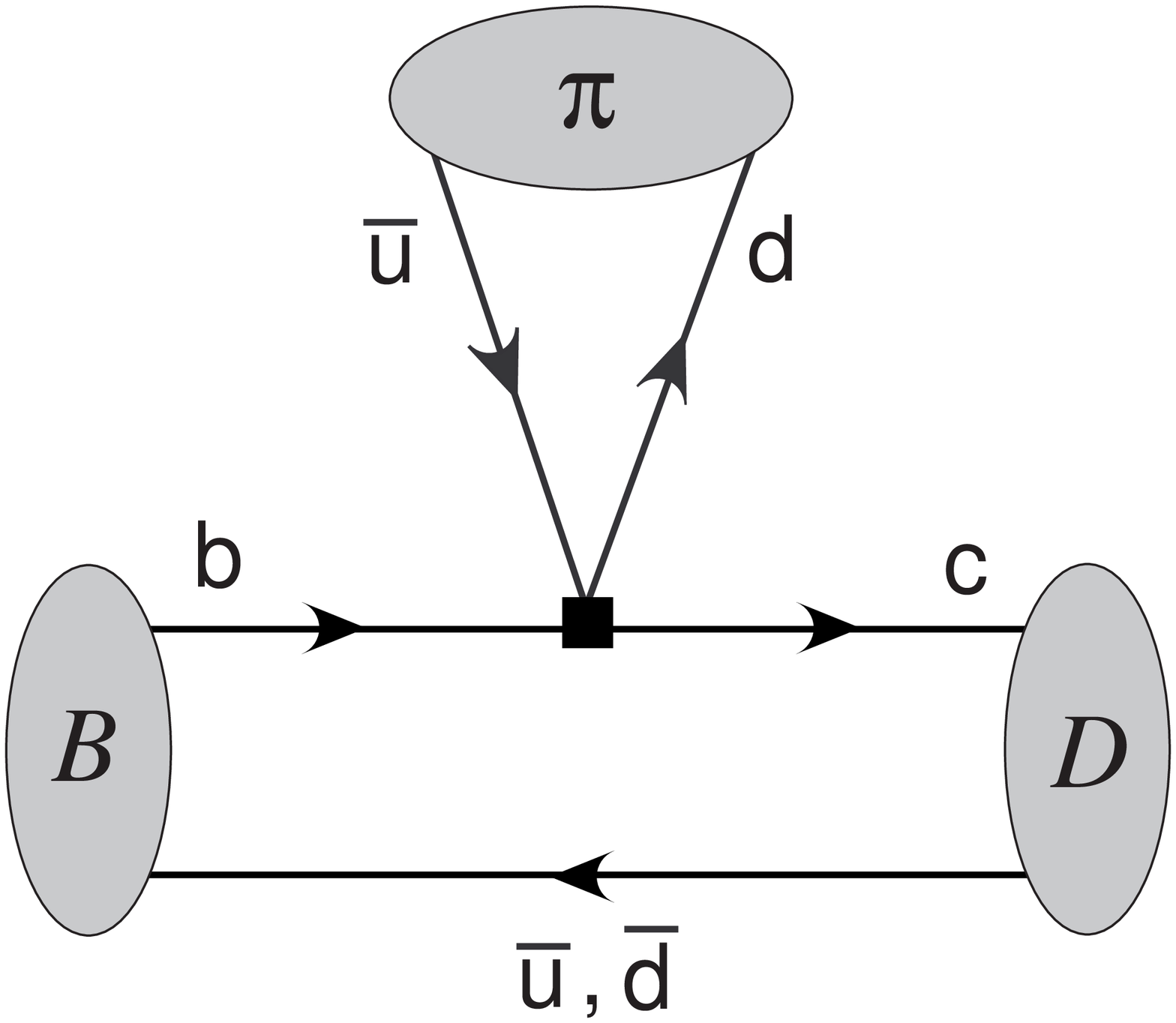}
  \includegraphics[height=.1\textheight]{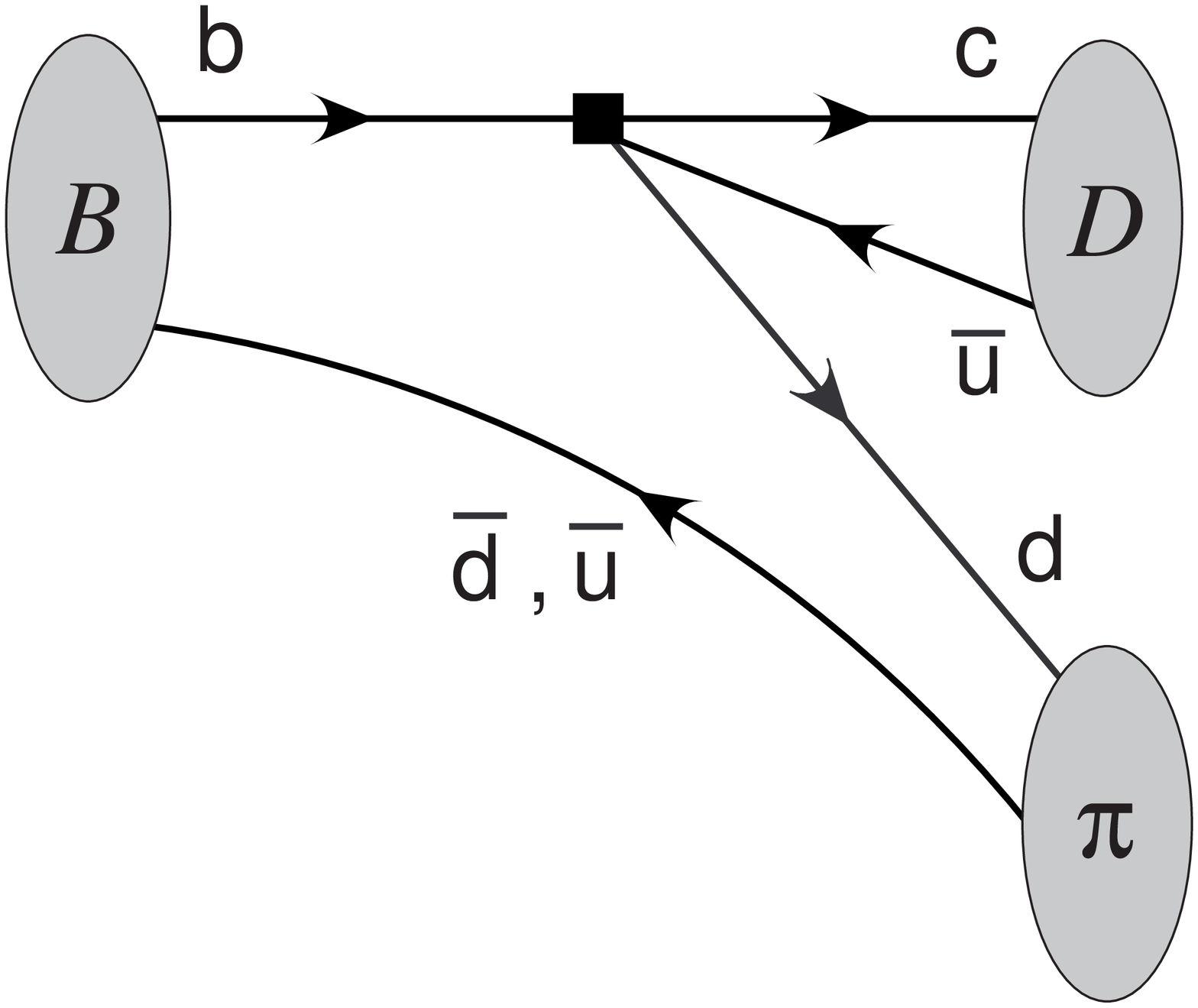}
  \includegraphics[height=.1\textheight]{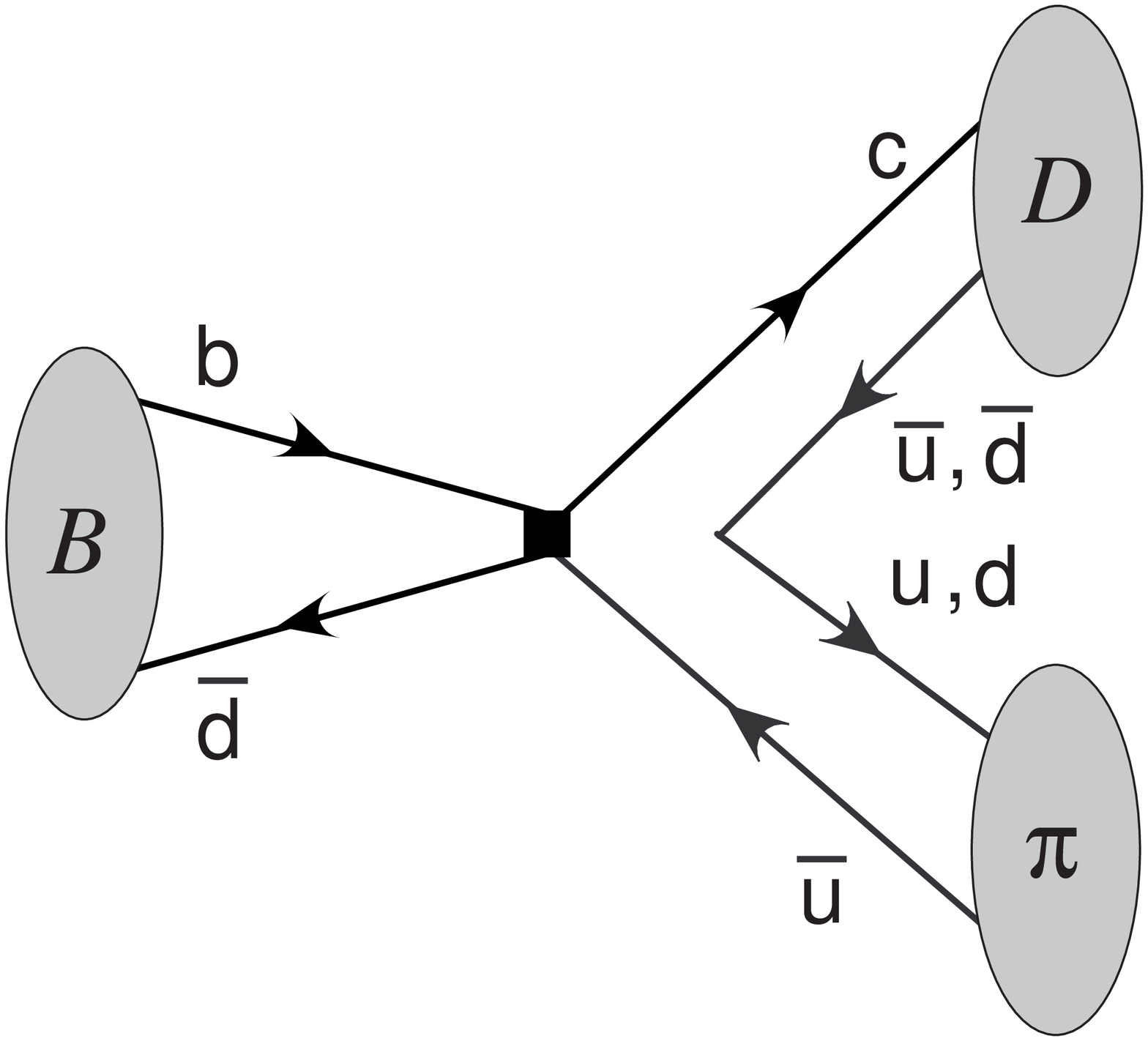}
  \caption{$B\to D\pi$ flavor topologies, T, C, and E respectively. For the
    color-suppressed decays  only C and E contribute.}
  \label{fig:cs}
\end{figure}

Color-suppressed decays were investigated in Ref.~\cite{Mantry:2003uz} using SCET.
For $B\to D\pi$ decays the four quark operators which contribute are
[$\Gamma=V-A$]
\begin{eqnarray} \label{Hw}
  H_W = \frac{G_F V_{cb} V_{ud}^*}{\sqrt{2}} \big[ C_1 (\bar c b)_{\Gamma}
      (\bar d u)_{\Gamma} + 
    C_2 (\bar c_i b_j)_{\Gamma} (\bar d_j u_i)_{\Gamma} \big] \,,
\end{eqnarray} 
with flavor contractions shown by the Fig.~\ref{fig:cs} diagrams.
For the amplitudes we use $A_{+-}=A(\bar B^0\to D^+\pi^-)$, $A_{0-}=A(B^-\to
D^0\pi^-)$, and $A_{00} = A(\bar B^0 \to D^0\pi^0)$. Written in terms of 
isospin amplitudes
\begin{eqnarray}
  A_{+-} &=& T+E = \frac{1}{\sqrt{3}} A_{3/2} + \sqrt{\frac23} A_{1/2} \,,\nn\\
  A_{0-} &=& T+C = \sqrt{3}\, A_{3/2} \,, \\
  A_{00} &=& \frac{C-E}{\sqrt{2}} 
     =\sqrt{ \frac23} A_{3/2} - \frac{1}{\sqrt3} A_{1/2} 
     \,. \nn
\end{eqnarray}
The amplitudes for decays to $B\to D^{(*)}\rho$ are defined in a similar fashion.

In the large $N_c$ limit $C/T\sim E/T\sim 1/N_c$ (where we take $C_1\sim 1$ and
$C_2\sim 1/N_c$). The color-allowed amplitudes $A_{+-}$ and $A_{0-}$ are
described by a factorization
theorem~\cite{Dugan:1991de,Politzer:1991au,Beneke:2000ry}, proven with
SCET~\cite{Bauer:2001cu}
\begin{eqnarray} \label{Alo}
  A^{(*)} = N^{(*)}\: \xi(w_{max}) \int_0^1\!\!dx\: 
   T^{(*)}(x,m_c/m_b)\: \phi_\pi(x) +\ldots ,
\end{eqnarray}
where $\xi(w_{max})$ is the Isgur-Wise function at maximum recoil, $\phi_\pi(x)$
is the light-cone distribution function for the pion, $T=1+ O(\alpha_s)$ is the
hard scattering kernel, and $N^{(*)}= \frac{G_F}{\sqrt{2}} V_{cb} V_{ud}^* E_\pi
f_\pi \sqrt{m_{D^{(*)}} m_B}(1+m_B/m_{D^{(*)}})$. The ellipses in
Eq.~(\ref{Alo}) denote terms suppressed by $\Lambda/Q$ where
$Q=\{m_b,m_c,E_\pi\}$. In the heavy quark limit, Eq.~(\ref{Alo}) predicts
$A=A^*$, so $Br(\bar B^0 \to D^+\pi^-)=Br(\bar B^0\to D^{*+}\pi^-)$ and
$Br(B^-\to D^0\pi^-)=Br(B^-\to D^{*0}\pi^-)$. This agrees well with the
experimental results~\cite{Ahmed:2002vm,Hagiwara:2002fs}, which yield
\begin{eqnarray} \label{LOp1}
  \frac{Br(\bar B^0 \to D^{*+}\pi^-)}{Br(\bar B^0 \to D^+\pi^-)} 
  &=& 1.03 \pm 0.14  \,,\nn\\
  \frac{Br(B^-\to D^{*0}\pi^-)}{Br(B^-\to D^0\pi^-)} 
  &=& 0.93 \pm 0.11  \,.
\end{eqnarray}
Eq.~(\ref{Alo}) also predicts $A_{+-}=A_{0-}$, however experimentally
$|A_{0-}/A_{+-}| = 0.77\pm 0.05$ for $D\pi$ and $0.81\pm 0.05$ for $D^*\pi$. We
will see that the reduction of these numbers from $1$ are explained by an SCET
power correction.  Other mechanisms for testing factorization for color-allowed
decays include using multibody states to make tests as a function of $q^2$ or
$w_{\rm max}$~\cite{Ligeti:2001dk}, looking for decays which do not occur in
naive-factorization~\cite{Diehl:2001xe}, or tests using inclusive $B\to
D^{(*)}X$ spectra or the equality of rates for particular multibody final states
$X$~\cite{Bauer:2002sh}.

\begin{figure}
  \includegraphics[height=.1\textheight]{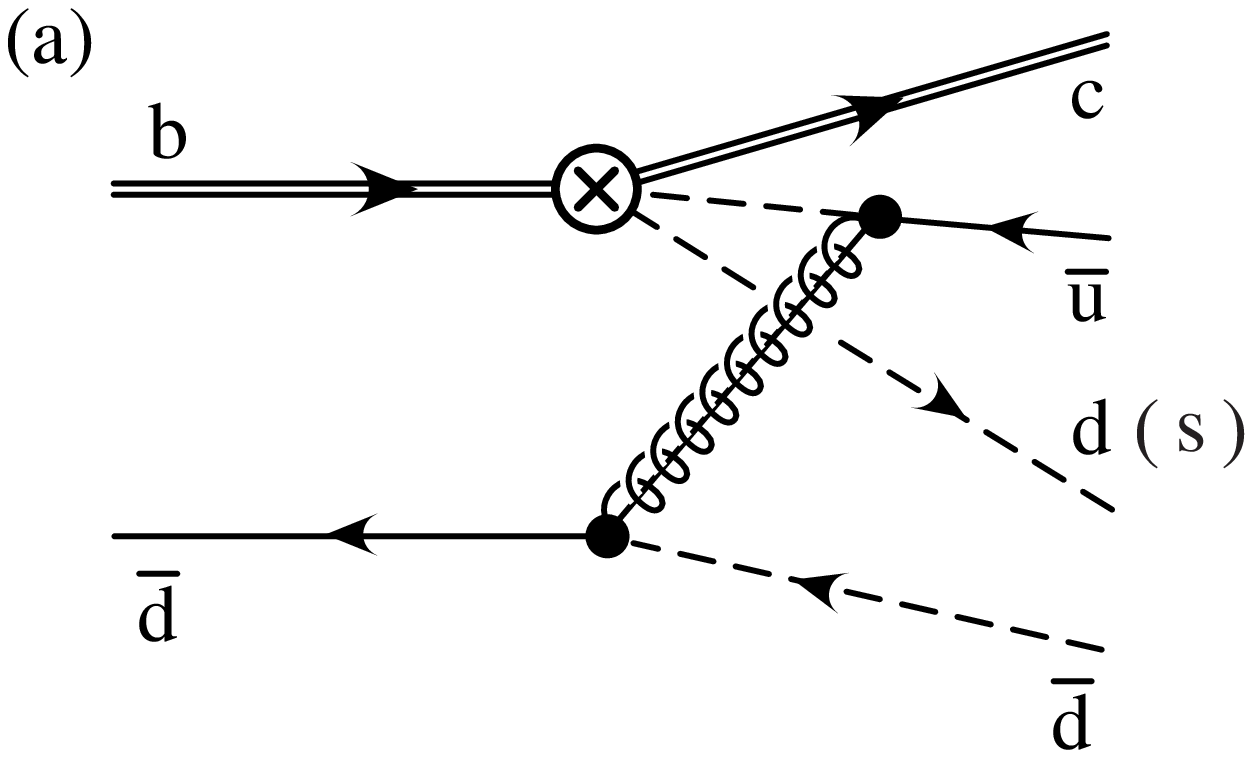}
  \includegraphics[height=.1\textheight]{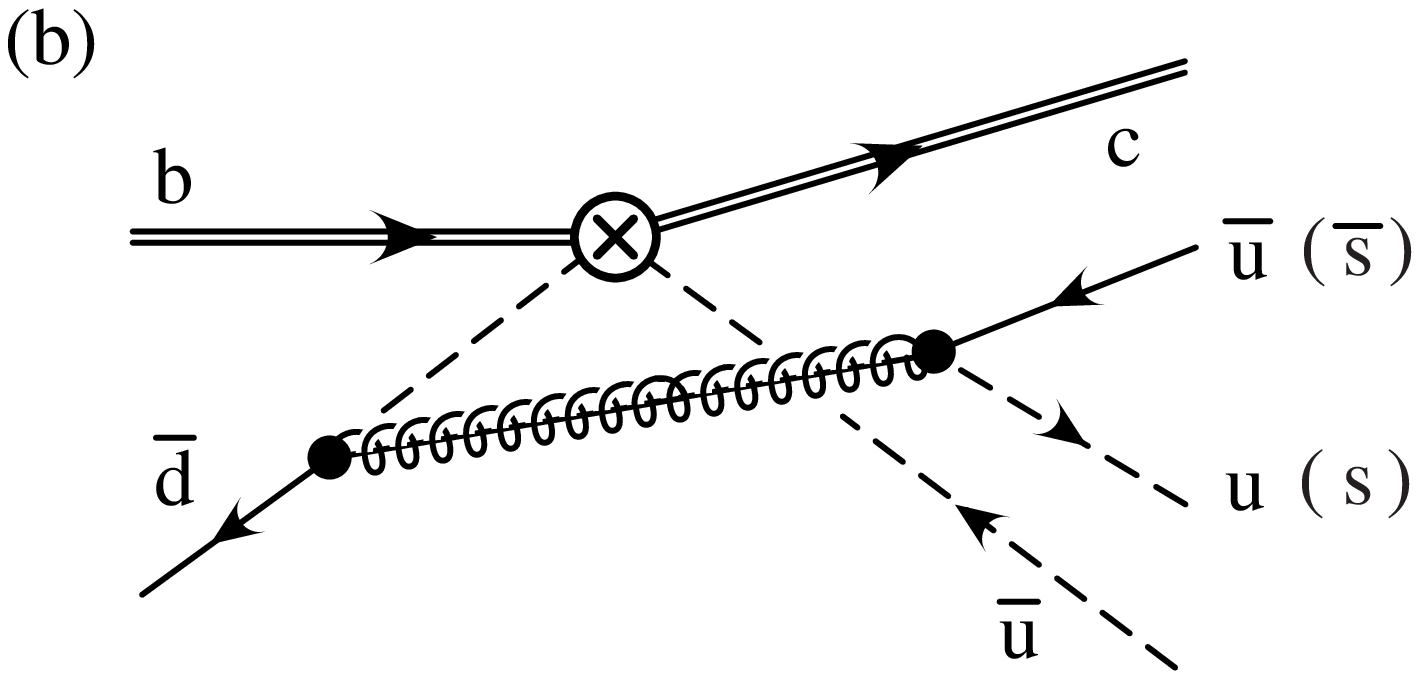}
    \caption{Diagrams in ${\rm SCET}_{\rm I}$ for tree level matching. 
      $\otimes$ denotes the operator $Q_j^{(0,8)}$ and the dots are insertions
      of ${\cal L}_{\xi q}^{(1)}$. The solid lines and double solid lines carry
      momenta $p^\mu\sim \Lambda$ and form the $B$ and $D$. The dashed lines are
      energetic collinear quarks that form the light meson $M$.}
  \label{fig:I}
\end{figure}

The color-suppressed amplitude $A_{00}$ has contributions from $C$ and $E$, but
not $T$.  With large $N_c$ very little can be said about the $C$ and $E$
contributions, besides the fact that we expect $A_{00} < A_{+-}\sim A_{0-}$.  In
SCET the amplitudes $C$ and $E$ are suppressed by $\Lambda/E_\pi$ relative to
$T$.  Despite this power suppression, predictive power is retained since only a
single type of ${\rm SCET}_{\rm I}$ time ordered product contributes to give the
proper quark rearrangement, $T( Q_{j}^{(0,8)}(0), i {\cal L}_{\xi q}^{(1)}(x), i
{\cal L}_{\xi q}^{(1)}(y))$~\cite{Mantry:2003uz}. This combination contributes
to both C and E as shown in Fig.~\ref{fig:I}.

When matched onto ${\rm SCET}_{\rm II}$ the time-ordered product gives a product
of soft $O_s^{(0,8)}$ and collinear $O_c$ operators. Thus $\langle D^{(*)}\pi |
O_s^{(0,8)} O_c | B\rangle = \langle D^{(*)} | O_s^{(0,8)} | B \rangle \langle
\pi| O_c | 0 \rangle$. The soft operator [$P_L=(1-\gamma_5)/2$]
\begin{eqnarray}
  O_s^{(0)} \!\!\!\! &=&\!\!\!\!
  (\bar h_{v'}^{(c)} S)  \nslash P_L \, (S^\dagger h_v^{(b)}) \:
  (\bar d\,S)_{k^+_1} \nslash P_L\, (S^\dagger u)_{k^+_2} 
  \,,
\end{eqnarray}
while $O_s^{(8)}$ is identical but with color structure $T^A\otimes T^A$. In
addition there are operators encoding ``long'' distance contributions in ${\rm
  SCET}_{\rm II}$ that are the same order in $\Lambda/Q$. These come from the
region of momentum space for Fig.~\ref{fig:I} where the gluon still has $p^2\sim
Q\Lambda$, but the quark propagator has $p^2\sim \Lambda^2$. 

\begin{figure}
  \includegraphics[height=.23\textheight]{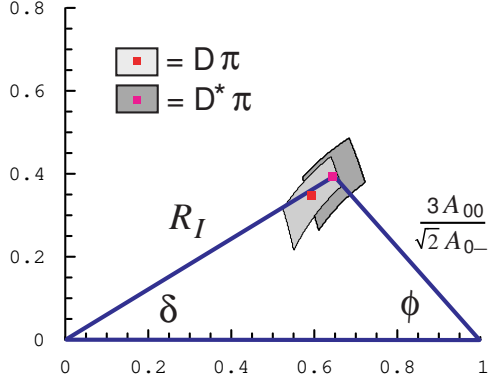}
  \caption{Experimental data for the $D\pi$ and $D^*\pi$ isospin
    triangles. This figure updates the one in Ref.~\cite{Mantry:2003uz} to include
  the recent BaBar data.}
  \label{fig:DD}
\end{figure}

Using heavy quark symmetry one can prove
\begin{eqnarray} \label{S}
  \langle D^{(*)} | O_s^{(0,8)} | B \rangle = S_L^{(0,8)}(k_1^+,k_2^+) \,,
\end{eqnarray}
so that the matrix elements for $\bar B^0\to D^0\pi^0$ and $\bar B^0\to
D^{*0}\pi^0$ are the same~\cite{Mantry:2003uz}. Furthermore, $S_L^{(0,8)}$ are
complex from their dependence on $n^\mu$, the {\em direction} of the light
meson, and encode a non-perturbative strong phase shift.  This leads to the
predictions
\begin{eqnarray} \label{p1}
  \delta(D^*\pi) &=& \delta(D\pi) \,,\\
  {Br(\bar B^0\to D^{*0}\pi^0)} &=& {Br(\bar B^0\to D^0\pi^0)} 
      \,,  \nn
\end{eqnarray}
where $\delta = {\rm arg}(A_{1/2} A_{3/2}^*)$ is the strong phase shift between
isospin amplitudes. The predictions in Eq.~(\ref{p1}) have corrections at
${O}(\alpha_s(Q))$ and $O(\Lambda/Q)$. For $M=\pi^0,\rho^0$ the long distance
amplitude is suppressed by $\alpha_s(Q)$. The current experimental
data~\cite{Coan:2001ei,Abe:2001zi,Aubert:2003sw} gives the world averages
[branching ratios below are in units of $10^{-3}$]
\begin{eqnarray} \label{dataDD}
  Br( D^{0}\pi^0) = 0.29\pm 0.03 \,,\ \  
   \delta(D\pi) = 30.4\pm 4.8^\circ , \\
  Br( D^{*0}\pi^0) = 0.26 \pm 0.05 \,, \
    \delta(D^*\pi) = 31.0\pm 5.0^\circ, \nn
\end{eqnarray}
showing good agreement with Eq.~(\ref{p1}).  If further data indicates agreement
of the angles beyond the current 17\% level then this would be an indication
that $\Lambda/Q$ corrections to Eq.~(\ref{p1}) are smaller than expected (or
perhaps absent), and the same applies for the ratios in Eq.~(\ref{LOp1}).  The
agreement can also be shown graphically. The isospin relation between amplitudes
implies that
\begin{eqnarray} \label{rel}
  1 = R_I + \frac{3}{\sqrt{2}} \frac{A_{00}}{A_{0-}} \,,
\end{eqnarray}
where $R_I = A_{1/2}/(\sqrt{2}A_{3/2}) = (A_{+-}-A_{00}/\sqrt{2})/A_{0-}$.
Eq.~(\ref{rel}) can be represented by a triangle in the complex plane. The
current world averages for $D\pi$ and $D^*\pi$ are shown in Fig.~\ref{fig:DD},
where the overlap of the $1$-$\sigma$ regions indicates the
agreement.

It is useful to note that Eq.~(\ref{p1}) provides a sensitive test of
SCET-factorization, and not just heavy quark symmetry. The basic reason is that
apriori ``soft'' gluon exchange between the $b$ or $c$ and the light quarks in the
pion spoils the prediction. To see this more clearly we can consider using just
HQET with full QCD for the light quarks. In this case the amplitude would be
\begin{eqnarray} \label{hqet}
 && \langle D^{(*)0}\pi^0 | (\bar h_{v'}^{(c)} \gamma^\mu P_L h_v^{(b)})
   (\bar d \gamma_\mu P_L u) | B^0 \rangle / \sqrt{m_B m_D} \nn \\ 
 && = {\rm Tr}\,[ \bar H_{v'}^{(c)} \gamma^\mu P_L H_v^{(b)} X_{\mu} ] \\
 &&\ +\frac{1}{m_c} {\rm Tr}\,[ \bar H_{v'}^{(c)} i\sigma^{\alpha\beta}
    \frac{1+{v'\!\!\!\!\!\!\slash}}{2} \gamma^\mu P_L H_v^{(b)}
    R_{\mu\alpha\beta} ] + \ldots
  \nn
\end{eqnarray}
where $H_v$ and $H_{v'}$ are HQET superfields and $X_\mu$ and
$R_{\mu\alpha\beta}$ are the most general tensor functions compatible with the
symmetries of QCD.  Here the $R_{\mu\alpha\beta}$ term has a chromomagnetic
operator insertion, $\bar h_{v'} \sigma_{\alpha\beta} G^{\alpha\beta} h_{v'}$ on
the charm quark.  Usually in HQET the $R_{\mu\alpha\beta}$ term would be
suppressed relative to the $X_\mu$ term. However, in Eq.~(\ref{hqet}) the pion
momentum $p_\pi^\mu = E_\pi n^\mu$ is an allowed four-vector in
$R_{\mu\alpha\beta}$.  Since $E_\pi/m_c \simeq 1.5$ the two terms are the same
size (and this will also be the case for all other terms in the $1/m_c$ heavy
quark expansion, ie. the expansion does not converge). Since $E_\pi\simeq
2.3\,{\rm GeV}$ the ``soft'' gluons are carrying hard momenta.  Terms like
$R_{\mu\alpha\beta}$ break the heavy quark spin symmetry and give $A(\bar B^0\to
D^{*0}\pi^0) \ne A(\bar B^0\to D^{0}\pi^0)$. In contrast, with SCET we can
expand in $\Lambda/E_\pi$ and factorize away the energetic pion. Thus the
matrix element in Eq.~(\ref{S}) has no $E_\pi$ dependence and is part of a
convergent expansion.

The SCET analysis also gives predictions for several channels where the data is
not yet available. For instance, the analysis above also applies for the $\rho$,
predicting
\begin{eqnarray} \label{p2}
  \delta(D^*\rho) &=& \delta(D\rho) \,,\\
  {Br(\bar B^0\to D^{*0}\rho^0)} &=& {Br(\bar B^0\to D^0\rho^0)} 
      \,.  \nn
\end{eqnarray}
A similar prediction can be made for decays to $D_s^{(*)}K^{(*)}$ except in this
case the long distance contributions to the amplitudes are not suppressed. This
means that both longitudinal and perpendicular polarizations occur at the same
order. The analog of Eq.~(\ref{p2}) is therefore:
\begin{eqnarray}\label{p2a}
  {Br(\bar B^0\to D^{*}_s K^- )} &=& {Br(\bar B^0\to D_s K^-)} 
      \,, \\
  {Br(\bar B^0\to D^{*}_s\bar K^{*-}_\parallel )} &=&
    {Br(\bar B^0\to D_s\bar K^{*-}_\parallel )}    \,, \nn
\end{eqnarray} 
where these color-suppressed decays are not part of an isospin triangle.
Cabbibo suppressed decays to kaons are more analogous to $D\pi$ and $D\rho$,
except that they also have long distance contributions which are not suppressed.
In this case the analog of Eq.~(\ref{p2}) is
\begin{eqnarray} \label{p3}
  \delta(D^*\bar K^0 ) &=& \delta(D\bar K^0  ) \,,\\
  {Br(\bar B^0\to D^{*0}\bar K^0 )} &=& {Br(\bar B^0\to D^0\bar K^0 )} 
      \,,  \nn\\
  \delta(D^*\bar K^{*0}_\parallel ) &=& \delta(D\bar K^{*0}_\parallel  ) 
      \nn \,,\\
  {Br(\bar B^0\to D^{*0}\bar K^{*0}_\parallel )} &=&
    {Br(\bar B^0\to D^0\bar K^{*0}_\parallel )}    \,.  \nn
\end{eqnarray}
The predictions in Eqs.~(\ref{p2}), (\ref{p2a}), and (\ref{p3}) will be tested
once data on $\bar B^0 \to D^{*0}\rho^0$, $\bar B^0 \to D^{*}_s K^{(*)-}$, and
$\bar B^0 \to D^{*0} \bar K^{(*)0}$ become available.  The significance of the
long distance terms will be tested by comparing $Br(\bar B^0\to
D^{*}_{s\parallel} \bar K^{*-}_\parallel )$ to $Br(\bar B^0\to D^{*}_{s\perp}
\bar K^{*-}_\perp )$ or $Br(\bar B^0\to D^{*0}_\parallel \bar K^{*0}_\parallel
)$ to $Br(\bar B^0\to D^{*0}_\perp \bar K^{*0}_\perp )$.

The full factorization theorem for color suppressed decays takes the
form~\cite{Mantry:2003uz}
\begin{eqnarray}
 A_{00}^{D^{(*)}M} \!\!\!\! &=& \!\!\!\!
   N_0^{M} \int\!\! dx\, dz\, d{k_1^+}\, d{k_2^+}\,
   T_{L\mp R}^{(i)}(z)\: J^{(i)}(z,x,k_1^+,k_2^+)\nn\\
 && \!\!\!\! \times S^{(i)}(k_1^+,k_2^+)\: \phi_M(x) + A_{long}^{D^{(*)}M}\,.
\end{eqnarray}
where $T_{L\mp R}^{(i)}$ are hard scattering kernels and $N_0^{M}= G_F
V_{cb}V_{ud}^* f_M \sqrt{m_B m_{D^{(*)}}}/2$. The non-perturbative dynamics is
contained in $\phi_M$, the light-cone distribution function for meson $M$, and
$S^{(i)}$, $i=0,8$, a generalized parton distribution function for the $B\to
D^{(*)}$ transition with $k_1^+$ and $k_2^+$ being momentum fractions of the
light spectator quarks. Finally, the jet function $J^{(i)}$ is sensitive to
physics at the $\mu^2\sim E_M\Lambda$ scale and is responsible for the quark
rearrangement.

The predictions discussed above are all valid independent of the form of
$J^{(i)}$, meaning to all orders in $\alpha_s(\mu_0)$ at the intermediate scale,
$\mu_0^2\sim E_M\Lambda$. If we expand $J$ in powers of $\alpha_s(\mu_0)$ then
this introduces additional uncertainty, but gives further predictions. At lowest
order we have
\begin{eqnarray} \label{p4}
  A_{00}^{D^{(*)}M} \!\!\!\! &=&\!\!\!\!
   N_0^M C_L^{(0)} \frac{16\pi\alpha_s(\mu_0)}{9} 
    s_{\rm eff}(\mu_0) \langle x^{-1} \rangle_M \,,
\end{eqnarray}
where $C_L^{(0)}= C_1+C_2/3$, $\langle x^{-1} \rangle_M = \int dx/x\:
\phi_M(x)$, and $s_{\rm eff}=-s^{(0)} + C_2/(4 C_1+4/3 C_2)\, s^{(8)}$ with
$s^{(0,8)}= \int dk_1^+ dk_2^+/(k_1^+ k_2^+)\: S^{(0,8)}(k_1^+,k_2^+)$.
Corrections to Eq.~(\ref{p4}) are $O(\alpha_s(\mu_0))$ and $O(\Lambda/Q)$. At
this order the strong phase $\phi$ in Fig.~\ref{fig:DD} is generated by $s_{\rm
  eff}$ and so is independent of whether $M=\pi$ or $M=\rho$.  Therefore, we
predict that $\phi$ is universal for $D^{(*)}\pi$ and $D^{(*)}\rho$.

For the ratio of charged amplitudes Eq.~(\ref{p4}) can be used to predict the
leading power correction,
\begin{eqnarray} \label{p5}
  R_c^{D^{(*)}M} &=& 
  \frac{A(\bar B^0\to D^{(*)+}M^-)}{A(B^-\to D^{(*)0}M^-)} \\
   &=& 1 - \frac{16\pi\alpha_s m_{D^{(*)}}}{9 (m_B + m_{D^{(*)}})}
     \frac{\langle x^{-1} \rangle_M}{\xi(\omega_0)}\ \frac{s_{\rm eff}}{E_M }\,. \nn
\end{eqnarray}
A value $s_{\rm eff}\simeq (430\,{\rm MeV}) e^{i\, 44^\circ}$ gives
$|R_c^{D\pi}| \simeq 0.8$, fitting the experimental values given below
Eq.~(\ref{LOp1}) with parameters of natural size.  In naive factorization the
correction term in $R_c$ would depend on the decay constant $f_M$, however for
the true factorization theorem that gives Eq.~(\ref{p5}) this turns out not to
be the case. The observed similarity between $R_c$ for $D\pi$ and $D\rho$ can be
explained by having $\langle x^{-1}\rangle_\pi \simeq \langle x^{-1}
\rangle_\rho$ and is not spoiled by the fact that $f_\rho/f_\pi\simeq 1.6$.
Experimentally~\cite{Hagiwara:2002fs}
\begin{eqnarray} 
   \frac{|R_c^{D\pi}|}{|R_c^{D\rho}|} &=&  0.96 \pm 0.13\,. 
\end{eqnarray}
With this approximate equality and the $\phi^\pi= \phi^\rho$ prediction, we
would expect that the strong phase $\delta^{D\rho}\simeq\delta^{D\pi}$, or in
other words that the $D\pi$ and $D\rho$ triangles (as in Fig.~\ref{fig:DD}) will
be similar.  If this turns out not to be the case then it would indicate that
there are substantial $\alpha_s^2(\mu_0)$ corrections to $J^{(0,8)}$. This would
mean that the subset of predictions that follow from Eq.~(\ref{p4}), which
depend on a perturbative expansion for $J^{(0,8)}$, should not be trusted.
Predictions for color-suppressed decays using other methods have been discussed
in
Refs.~\cite{Neubert:2001sj,Chiang:2002tv,Keum:2003js,Chua:2001br,Wolfenstein:2003pc,Calderon:2003vc}.

\section{Baryon Decays}

Recently, the authors in Ref.~\cite{Leibovich:2003tw} have used SCET to make
model-independent factorization predictions for baryon decays.  The main results
for $\Lambda_b\to \Lambda_c \pi$, $\Lambda_c \rho$, $\Sigma_c^{(*)} \pi$,
$\Sigma_c^{(*)} \rho$, and $\Xi_c^{(\prime,*)}K$ are briefly summarized here.
The notation is $\Sigma_c=\Sigma_c(2455)$ and $\Sigma_c^*=\Sigma_c(2520)$.

\begin{figure}[t]
 \begin{tabular}{c}
 { \includegraphics[width=.15\textwidth]{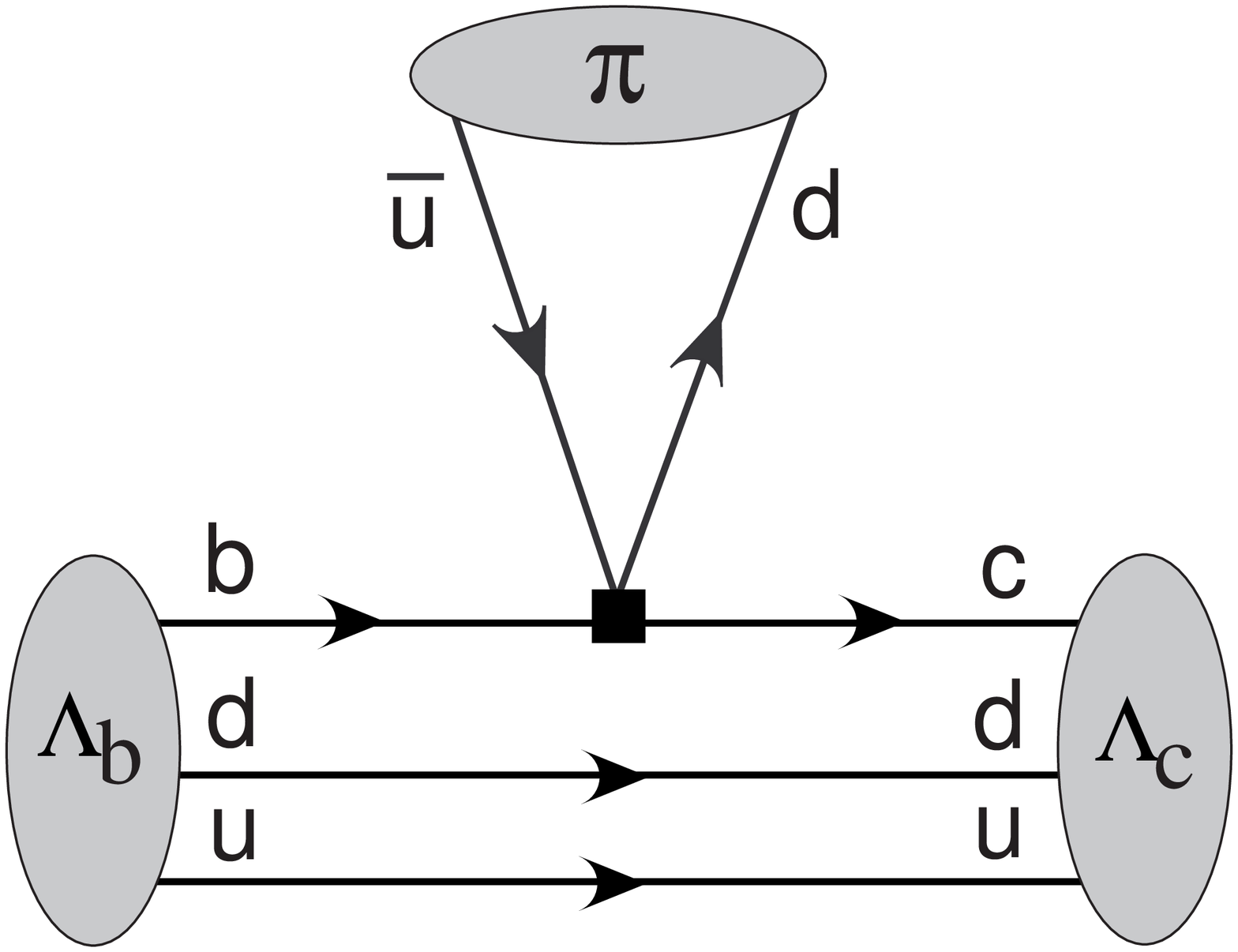} \hspace{0.1cm}
  \includegraphics[width=.15\textwidth]{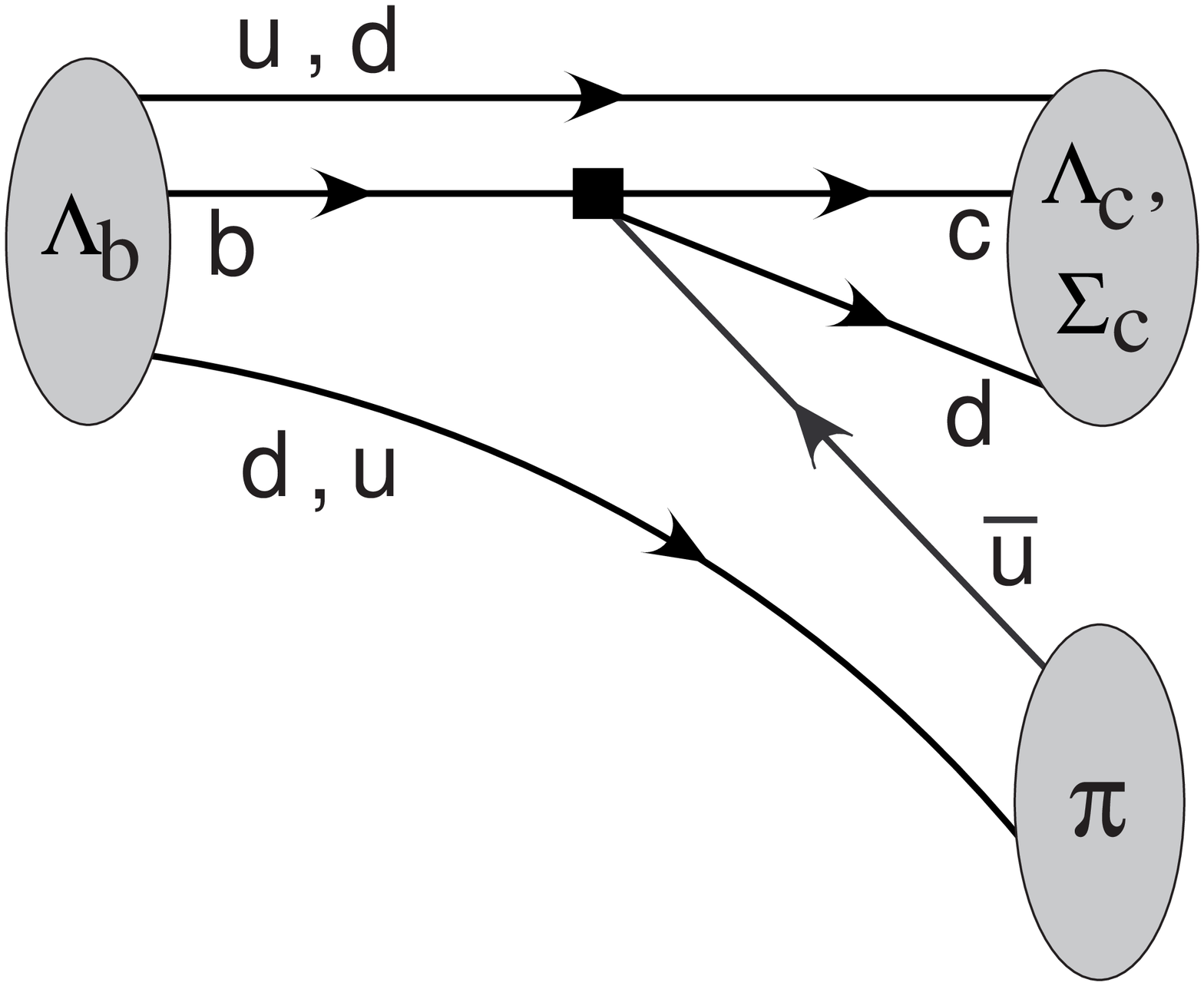} }
   \\
  {\includegraphics[width=.15\textwidth]{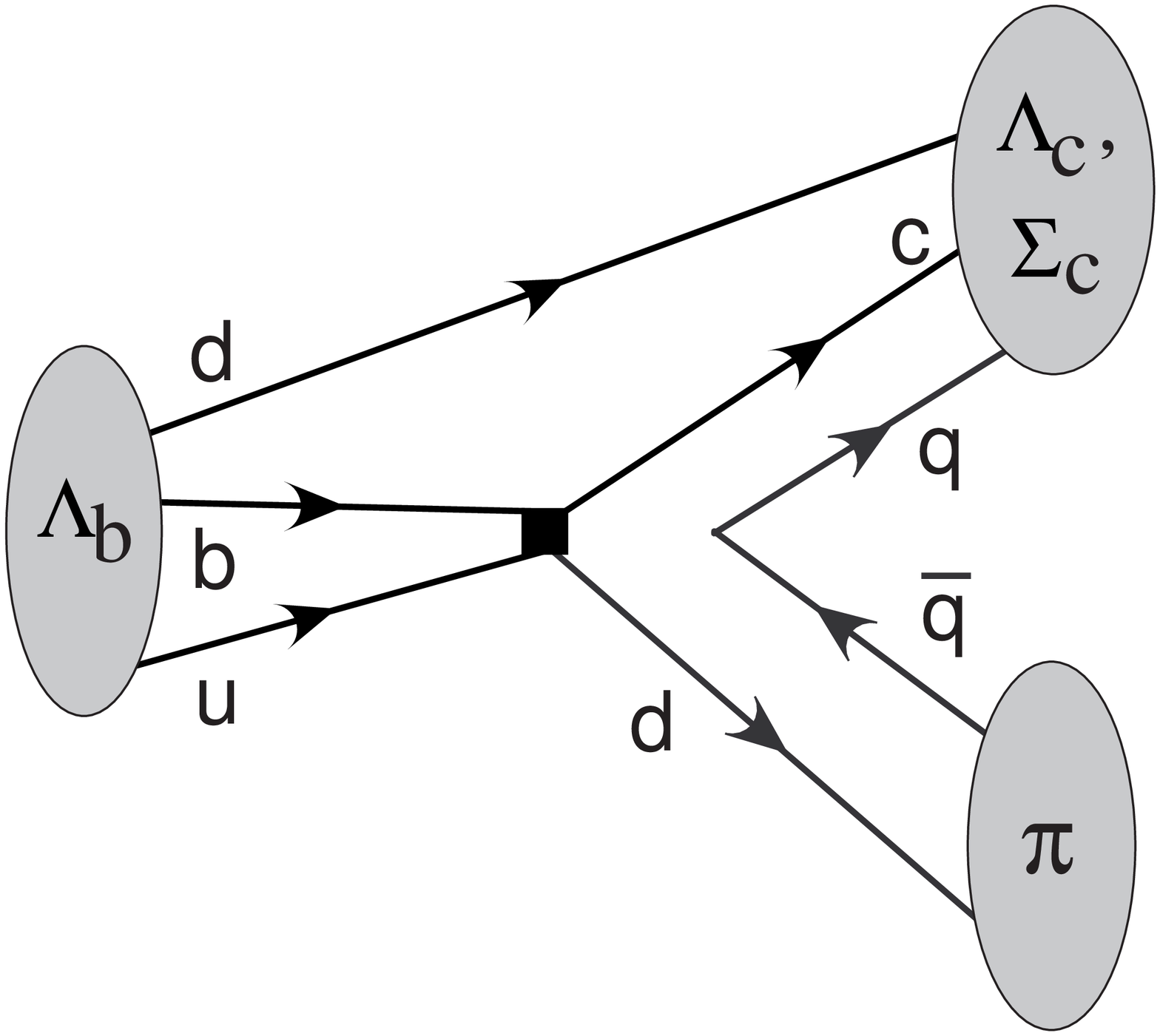} \hspace{0.1cm}
  \includegraphics[width=.15\textwidth]{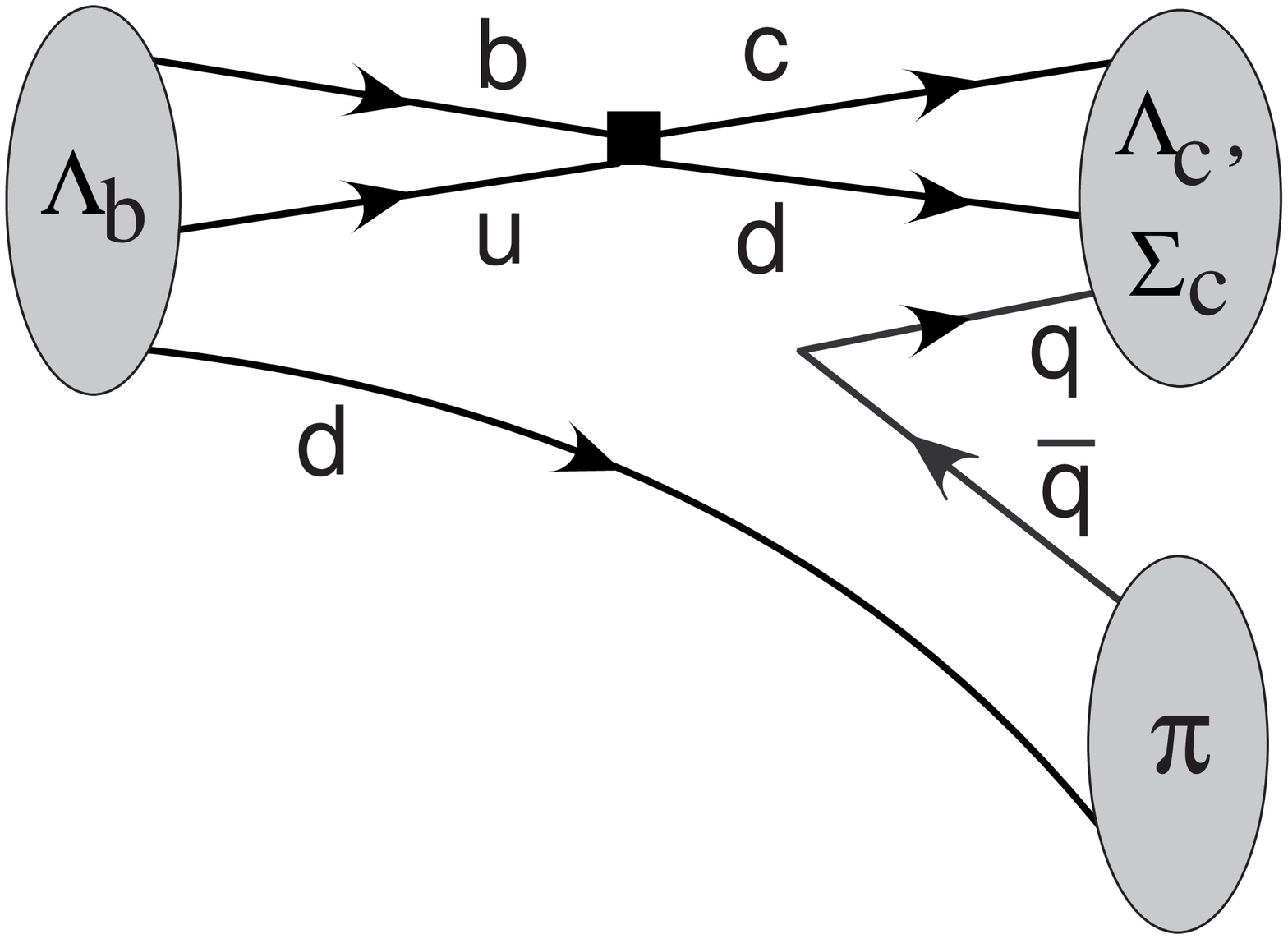} }
 \end{tabular}
\vspace{-0.25cm}
\caption{Classes of diagrams for $\Lambda_b$ decays, giving amplitudes $T$
  (tree) and $C$ (color-commensurate) for the top two diagrams, and $E$
  (exchange), and $B$ (bow-tie) for the bottom two. Bow-tie diagrams are unique
  to baryon decays~\cite{Leibovich:2003tw}.}
\label{fig:bdiagrams}
\end{figure}

The electroweak Hamiltonian for these baryon decays is in Eq.~(\ref{Hw}). The
diagrams for the flavor contractions differ from the meson decays and are shown
in Fig.~\ref{fig:bdiagrams}. The decays to $\Lambda_c$ get contributions from
$T$, $C$, $E$, and $B$, decays to $\Sigma_c^{(*)}$ have contributions from $C$,
$E$, and $B$, and decays to $\Xi_c$ only from the $E$ and $B$ amplitudes.

In the large $N_c$ limit $C/T\sim E/T\sim N_c^0$, while $B/T \sim N_c$. Thus,
even $\Lambda_b\to \Lambda_c\pi$ decays do not factorize in the large $N_c$
limit. The extra factors of $N_c$ arise from the choice of which of the $N_c$
quarks in the $\Lambda_{b}$ and/or $\Lambda_c$ participate in the weak
interaction.

Expanding in $\Lambda/Q$ where $Q=\{m_b,m_c,E_\pi\}$ using SCET one finds
$C/T\sim E/T\sim \Lambda/Q$ and $B/T\,\lesssim\, \Lambda^2/Q^2$. For $\Lambda_b\to
\Lambda_c\pi$ the leading order result is from $T$ and gives
\begin{eqnarray} \label{Alob}
  A^{\Lambda_c\pi}\!\!\!\! &=&\!\!\!\! N_{L}\: \zeta(w_{max}) \int_0^1\!\!dx\: 
   T_{L}(x,m_c/m_b)\: \phi_\pi(x) \\
  &+&\!\!\!\! N_{R}\: \zeta(w_{max}) \int_0^1\!\!dx\: 
   T_{R}(x,m_c/m_b)\: \phi_\pi(x)
   +\ldots \nn ,
\end{eqnarray}
where $\zeta(w_{max})$ is the $\Lambda_b\to \Lambda_c$ Isgur-Wise function at
maximum recoil, $T_{L,R}$ are hard scattering kernels, and $N_{L,R} = {\sqrt{2}}
{G_F}V_{cb} V_{ud}^* E_\pi f_\pi \sqrt{m_{\Lambda_c} m_{\Lambda_b}}\ \bar u(v')
\nslash P_{L,R}\, u(v)$ with $\bar u(v) u(v)=2$ and the states normalized as in
the PDG. The other factors are the analogs of those in Eq.~(\ref{Alo}). The
value of $\zeta(w_{max})$ will be determined by $q^2$ spectrum measurements of
$\Lambda_b\to \Lambda_c \ell\bar\nu_\ell$ which are not yet available. If
$\zeta(w_{max})$ is similar to $\xi(w_{max})$ then Eq.~(\ref{Alob}) predicts
that $Br(\Lambda_b\to \Lambda_c\pi) \sim 2 Br(\bar B^0 \to D^+\pi^-)$ in
agreement with the measurements from CDF~\cite{CDF:6396}.

For baryons the analog of the color-suppressed decays $\bar B^0\to
D^{(*)0}\pi^0$ are $\Lambda_b\to \Sigma_c^{(*)0}\pi^0$ and its isospin partner
$\Lambda_b\to \Sigma_c^{(*)+}\pi^-$. The leading amplitudes are $C\sim E$, while
$B$ is suppressed by an additional $\Lambda/Q$. Using heavy quark symmetry on
$\Lambda_b\to \Sigma_c^{(*)}$ matrix elements of the SCET operators
$O_s^{(0,8)}$ gives~\cite{Leibovich:2003tw}
\begin{eqnarray} \label{p2b}
  \frac{Br(\Lambda_b \to \Sigma^*_c \pi)}{Br(\Lambda_b \to \Sigma_c \pi)} = 2 \,,
\end{eqnarray}
up to corrections suppressed by $\Lambda/Q$ or $\alpha_s(Q)$. Here
$\Sigma_c^{(*)}\pi=\Sigma_c^{(*)0}\pi^0$ or $\Sigma_c^{(*)+}\pi^-$. A similar
prediction is also made for decays to a $\rho$,
\begin{eqnarray} \label{p3b}
  \frac{Br(\Lambda_b \to \Sigma^*_c \rho)}{Br(\Lambda_b \to \Sigma_c \rho)} = 2 \,.
\end{eqnarray}
Using $\Sigma_c^{(*)0}\rho^0\to \Lambda_c \pi^- \pi^+ \pi^-$ Eq.~(\ref{p3b}) may
be easier to test experimentally than Eq.~(\ref{p2b}).  For decays involving
cascades there can be sizeable long distance contributions, but we still expect
\begin{eqnarray}
  \frac{Br(\Lambda_b \to \Xi^*_c K)}{Br(\Lambda_b \to \Xi_c^\prime K)} &=& 2
  \,,\nn\\
  \frac{Br(\Lambda_b \to \Xi^*_c K^*_\parallel)}
    {Br(\Lambda_b \to \Xi_c^\prime K^*_\parallel)} &=& 2 \,.
\end{eqnarray}
The $Br(\Lambda_b \to \Xi_c K)$ is also expected to be of the same order of
magnitude since it occurs at this order in the power counting.

\section{Conclusion}
 
In this talk we reviewed the SCET predictions for non-leptonic decays with
charmed hadrons in the final state~\cite{Mantry:2003uz,Leibovich:2003tw}.  This
included the decays $\bar B^0\to D^{(*)+} \pi^-$, $B^-\to D^{(*)0}\pi^-$, and
$\Lambda_b\to \Lambda_c\pi$ which occur at leading order, as well as decays
which are power suppressed, $\bar B^0\to D^{(*)0} \pi^0$ and $\Lambda_b \to
\Sigma_c^{(*)}\pi$. Analogous decays where the $\pi$ is replaced by a $\rho$ or
kaon were also discussed.  For $\bar B^0\to D^{(*)0} \pi^0$ we updated the
experimental comparison in Fig.~\ref{fig:DD} and Eq.~(\ref{dataDD}) to take into
account the new BaBar results~\cite{Aubert:2003sw}.

\vspace{-0.4cm}

\begin{theacknowledgments}
  I.S. would like to thank A.~Leibovich, Z.~Ligeti, and M.~Wise for
  collaboration on the baryon results discussed here.  This work was supported
  in part by the Department of Energy under the cooperative research agreement
  DF-FC02-94ER40818. I.S. was also supported by a DOE Outstanding Junior
  Investigator award.
\end{theacknowledgments}
\vspace{-0.3cm}


\bibliographystyle{aipproc}   

\bibliography{iain_proc}

\IfFileExists{\jobname.bbl}{}
 {\typeout{}
  \typeout{******************************************}
  \typeout{** Please run "bibtex \jobname" to optain}
  \typeout{** the bibliography and then re-run LaTeX}
  \typeout{** twice to fix the references!}
  \typeout{******************************************}
  \typeout{}
 }

\end{document}